\documentclass[preprint,12pt,longnamesfirst]{aastex}
\usepackage{latexsym}

\shorttitle{Protostars in N\,51D}
\shortauthors{}

\begin{document}

\title{Protostars, Dust Globules, and a Herbig-Haro Object in 
the LMC Superbubble N~51D}

\author{Y.-H. Chu\altaffilmark{1}, 
        R.~A.\ Gruendl\altaffilmark{1}, 
        C.-H.\ R.\ Chen\altaffilmark{1},
        B.~A.\ Whitney\altaffilmark{2},
        K.~D.\ Gordon\altaffilmark{3},
        L.~W.\ Looney\altaffilmark{1},
        G.~C.\ Clayton\altaffilmark{4},
        J.~R.\ Dickel\altaffilmark{1},
        B.~C.\ Dunne\altaffilmark{1},
        S.~D. Points\altaffilmark{5},
        R.~C.\ Smith\altaffilmark{5}, and
        R.~M.\ Williams\altaffilmark{1}
}
\altaffiltext{1}{Astronomy Department, University of Illinois, 
        1002 W. Green Street, Urbana, IL 61801}
\altaffiltext{2}{Space Science Institute, 4750 Walnut Street, Suite 205, 
                 Boulder, CO 80301.}
\altaffiltext{3}{Steward Observatory, University of Arizona, Tucson, AZ 85721}
\altaffiltext{4}{Department of Physics and Astronomy, Louisiana State University,
      Baton Rouge, LA 70803}
\altaffiltext{5}{Cerro Tololo Inter-American Observatory,
      Casilla 603, La Serena, Chile}

\begin{abstract}
Using {\it Spitzer Space Telescope} and {\it Hubble Space Telescope} 
observations of the superbubble N51D, we have identified three 
young stellar objects (YSOs) in dust globules, and made the first 
detection of a Herbig-Haro object outside the Galaxy.
The spectral energy distributions of these YSOs suggest young massive 
stars with disk, envelope, and outflow cavities.  The interstellar 
conditions are used to assess whether the star formation was spontaneous
or induced by external pressure. 

\end{abstract}


\section{Introduction}

The {\it Spitzer Space Telescope}, with its high angular resolution 
and sensitivity in the mid- and far-infrared wavelengths, opens a new
window to view the formation of individual massive stars not only in 
the Galaxy but also in nearby galaxies, such as the Large Magellanic 
Cloud (LMC).
Using {\it Spitzer} observations, \citet{Jetal05} reported four young 
stellar objects (YSOs) in the \ion{H}{2} complex N\,159 in the LMC.
We have obtained {\it Spitzer} observation of the superbubble
LH$\alpha$\,120-N\,51D \citep[designation from][]{Henize56}, 
or N\,51D for short, in the LMC, and are surprised to find
YSOs projected within the superbubble where the density is expected 
to be low.

N\,51D has been partially imaged with the {\it Hubble Space Telescope 
(HST)} in the H$\alpha$ and [\ion{S}{2}] lines \citep{Cetal00}.
These {\it HST} images allow us to identify dust globules and
Herbig-Haro (HH) objects associated with YSOs.
We have modeled the spectral energy distributions (SEDs) of these 
YSOs in dust globules to estimate their physical parameters.   
We have further used H$\alpha$ and X-ray observations to 
determine the external gas pressure of the dust globules, in order
to assess whether the star formation was spontaneous or induced.
This {\it Letter} reports our analysis of {\it Spitzer} and {\it HST} 
observations of the three YSOs embedded in dust globules in N\,51D.

\section{Observations and Data Reduction}

Our {\it Spitzer} observations of N\,51D were made with the 
Infrared Array Camera \citep[IRAC;][]{Fetal04} on 2004 December
16 and the Multiband Imaging Photometer for {\it Spitzer} 
\citep[MIPS;][]{Retal04} on 2005 April 3.  
The IRAC observations were obtained using the mapping mode to cover 
a $\sim$30\arcmin$\times$20\arcmin\ area in the 3.6, 4.5, 5.8, and 
8.0 $\mu$m bands.
The four band maps overlap in a smaller region, 
$\sim$23\arcmin$\times$20\arcmin\ in size.
Each position in the map was observed with five 30 s integrations 
taken in the high dynamic range mode and using medium dithers in a 
cyclic pattern between exposures to aid in the rejection of transients.  
The resulting integration time at each location in the image was 
$\sim$150 s.
Our analyses use the post-BCD (Basic Calibrated Data) products 
resulting from standard pipeline processing (version S11.0.2).  

The MIPS observations were obtained using the scan map mode
in the 24, 70, and 160 $\mu$m bands.  
The medium scan rate was used to map a region comprised of 
sixteen 0$\rlap{.}{^\circ}$5 scan legs with a cross-scan step of 
148\arcsec\ to cover a region 30\arcmin$\times$20\arcmin\ in all 
three MIPS bands.  
The MIPS DAT v3.00 \citep{Getal05} was used to do the basic
processing and final mosaicking of the individual images.  
In addition, extra processing steps on each image were carried
out.  At 24~$\mu$m, these include readout offset correction and 
division by a scan-mirror-independent flat field.
At 70 and 160~$\mu$m the extra step was a pixel-dependent 
background subtraction for each map (using a low-order polynomial 
fit to the source free regions). 
The final mosaics have exposure times of roughly 200, 80, and 
18 s pixel$^{-1}$ for 24, 70, and 160~$\mu$m, respectively.
The MIPS photometry was determined using the point source fitting
software StarFinder \citep{Detal00} and a smoothed STinyTim model
point spread function.


\section{YSOs in the Superbubble N\,51D}

\subsection{Identification of YSOs}

YSOs can be distinguished from normal stars because of their
excess infrared emission from circumstellar dust.
To show the locations of YSOs and normal stars with respect to 
the ionized interstellar gas, we present in Figure 1a a color
composite of IRAC 8 $\mu$m (red), optical continuum at 5000 \AA\ 
(blue), and H$\alpha$ (green) images.
In this color image, the ionized interstellar gas appears green
and diffuse, most normal stars appear blue and unresolved, YSOs 
appear red and unresolved, and the interstellar dust emission
appears red and diffuse.

The ionized gas shell of the superbubble N\,51D is elongated along 
the N-S direction with a faint extension to the southwest.
The 8 $\mu$m diffuse dust emission is the brightest in filaments 
along the periphery of the ionized gas shell; it also shows a 
faint component extending across the entire field.
Two OB associations, LH\,51 and LH\,54 \citep{LH70,OS98}, exist
within the superbubble, appearing as concentrations of blue stars
near the western and eastern shell rims in Figure 1a.
Bright YSOs are seen within the boundary of the N\,51D superbubble.
Some YSOs along the north and west shell rims are associated with 
bright filamentary dust emission, while the other YSOs appear 
isolated.
Some YSOs are projected within the OB associations LH\,51 and 
LH\,54, but the YSOs on the north rim are not near either 
association.

Note however that not all red stars in Figure 1a are YSOs.
These red stars may appear red or blue in the color composite 
of IRAC 3.6, 4.5, and 8.0 $\mu$m images displayed in Figure 1b.
To determine the nature of these stars, we examine their
colors quantitatively and present an IRAC color-color diagram 
([3.6]$-$[4.5] versus [5.8]$-$[8.0]) and a color-magnitude 
diagram ([8.0] versus [3.6]$-$[8.0]) of N\,51D in Figure 2.
Similar diagrams presented by \citet{Aetal04} and 
\citet{Wetal04} for star forming regions show that
normal stars have zero colors and that YSOs exhibit a 
range of red colors depending on their classes, or 
amounts of circumstellar material.
For the color schemes we use, YSOs appear red in both Figures 
1a and 1b, but normal stars appear blue in Figure 1b even if 
they appear red in Figure 1a.



We use the objects 1-4 marked in Figures 2 and 3 as examples.
Objects 1-3 are YSOs and they appear red in both Figures 1a and 1b.  
Object 4 has been cataloged as L54S-81 with $V$ = 16.82$\pm$0.01,
$(B-V)$ = 1.67$\pm$0.05 \citep{Oey96}, and $K$ = 11.64$\pm$0.02
(2MASS).
Its colors and magnitudes are consistent with a Galactic M4\,V 
star at a distance of 67 pc; thus star 4 appears blue in 
Figure 1b and shows nearly zero colors in Figure 2a.

\subsection{Dust Globules and Herbig-Haro Object}

The eastern rim of the N\,51D superbubble has been imaged
by the {\it HST} WFPC2 in the H$\alpha$ and [\ion{S}{2}]
$\lambda\lambda$6716, 6731 lines \citep{Cetal00}.
Figure 3a shows a color composite of the {\it HST} H$\alpha$ (green)
and [\ion{S}{2}] (blue) images and {\it Spitzer} 8 $\mu$m image (red).
While the foreground M dwarf (object 4 in Fig.\ 2) appears isolated,
the three YSOs (objects 1--3 in Fig.\ 2) are each associated with 
a small dust feature whose surface gas is ionized.

Figure 3b shows a close-up H$\alpha$ image of the dust globules 
associated with YSO-1 and YSO-2.
The dust globules are each about 5$''$ in size, corresponding
to 1.25 pc, slightly larger than typical Bok globules in 
the Galaxy and in the LMC \citep{Cletal91,Getal99}.
The bright star to the east of YSO-2 is the WR star HD\,36402
\citep{BAT99}.

The dust feature associated with YSO-3, shown in Figure 3c,
has a different morphology.
Its surface toward the center of the N\,51D superbubble is
photoionized, but its back side is not.
The [\ion{S}{2}] close-up in Figure 3d shows additional
knots of emission extending from the north rim outward
along an almost straight line.
These knots are faint in H$\alpha$.
Their [\ion{S}{2}]/H$\alpha$ ratios increase outward from 
0.5 to 0.65, significantly higher than those of the 
photoionized surface of the dust feature, 0.25$\pm$0.05.
The alignment of these knots and their high [\ion{S}{2}]/H$\alpha$
ratios are characteristics that are frequently seen in HH objects 
\citep{RB01}; thus we suggest that {\it this is first HH object 
detected in the LMC}.


\subsection{Physical Properties of the YSOs}

We have measured the flux densities of the three YSOs in N\,51D in all
IRAC and MIPS bands, and listed them in Table 1.  
The SEDs were modeled with a 2-D radiation transfer code appropriate 
for YSOs \citep{Wetal03}.   These models include thermal dust 
emission but not PAH emission.  The circumstellar geometry consists 
of any combination of a rotationally flattened envelope, flared disk,
and bipolar cavities.
YSO-1 is fit with a source luminosity of 10,500 $L_\sun$ appropriate 
for a B2--B3 main sequence star, an envelope infall rate of 
$2 \times 10^{-4} M_\sun$~yr$^{-1}$, disk mass of 0.1 $M_\sun$,
and bipolar outflow cavity opening angle of 15\arcdeg.  
The disk and envelope centrifugal radius is 300 AU.   
These model parameters suggest a fairly young protostar.
The disk radius could be varied by 50\% and, with minor variations 
in other parameters, result in a similar SED.
The total envelope mass is $\sim 700 M_\sun$, but this is somewhat
dependent on our choice of outer radius and the radial density power 
law.
The disk mass is also uncertain since the long-wavelength contribution 
blends with the envelope.
However, without a disk, the envelope would have a steeper density 
profile and less near- and mid-IR flux density.
Our best model SED for YSO-1 is shown in Figure 4.
Ten inclinations are shown, and the best-fit 
inclination is $\sim50$\arcdeg.

YSO-2 is fit by a similar model as YSO-1: an illuminating B3 star 
(2500 $L_\sun$), an envelope mass of $\sim$210 $M_\sun$ and infall 
rate of $4 \times 10^{-4} M_\sun$~yr$^{-1}$, disk mass of 
0.1 $M_\sun$ and radius of 300 AU, and bipolar outflow cavity 
opening angle of 30\arcdeg.
The best-fit inclination is $\sim60$\arcdeg.

YSO-3 is interesting because it requires a massive envelope to fit 
the long-wavelength SED, but a nearly unextinguished line of sight 
to the illuminating central source, suggested by the optical data 
available for this source \citep[see Fig.~4;][]{Zetal02}.
We fit this with a $\sim$B4 star (1200 $L_\sun$), envelope mass
of $\sim$230 $M_\sun$ and infall rate of $5 \times 10^{-4} M_\sun$ 
yr$^{-1}$, disk mass of 0.2 $M_\sun$ and radius 400 AU, and bipolar 
cavity opening angle of 40\arcdeg.  
The best fit inclination is $\sim40$\arcdeg. 
This suggests either an evolved protostar with a substantial 
envelope and a large cavity carved out by outflows; or  
more than one source in the beam, an unextinguished source 
surrounded by a disk, and an embedded source with sufficient 
envelope mass to fit the long-wavelength SED.   
In either case, the model requires substantial warm dust, such 
as from a disk, to fit the mid-IR flux density.  
In the latter case, the HH object could come from either source.

In all of these sources, PAH emission in the [3.6], [4.5], and 
[8.0] IRAC bands is likely to contribute additional flux 
densities over the thermal emission accounted for by the models.
This is especially true in YSO-3 which is less embedded than 
the other two sources.
We have thus excluded these bands in estimating the best-fit 
inclination.


\section{Spontaneous or Induced Star Formation?}

The three dust globules that harbor YSOs in N\,51D are 
exposed to the radiation field of OB associations, similar
to the dust pillars in the Eagle Nebula \citep{Hetal96}.
The interstellar medium in N\,51D has been well-studied,
so we can use the external conditions of the dust globules
to assess whether the star formation was induced by 
external pressure.

The interior of a superbubble is expected to be filled with
shock-heated gas at 10$^6$--10$^7$ K \citep{Cetal95}.
The existence of hot gas in N\,51D has been confirmed by the 
diffuse X-ray emission detected by {\it Einstein} and {\it ROSAT} 
\citep{CM90,Detal01}.
Recent {\it XMM-Newton} observations provide high-quality X-ray 
spectra, and the best-fit thermal plasma emission models 
indicate that N\,51D's interior hot gas has an electron 
temperature of $\sim2.6\times10^6$ K and an electron density 
of $\sim0.03$ cm$^{-3}$ \citep{Cetal04}.
The thermal pressure of this hot gas is $P/k \sim 1.5\times10^5$
cm$^{-3}$ K, assuming a He to H number ratio of 1:10.

The ionized gas enveloping the dust globule around YSO-1 
has a peak H$\alpha$ surface brightness of
$1.4\times10^{-14}$ ergs~s$^{-1}$~cm$^{-2}$~arcsec$^{-2}$
along its limb-brightened rim, corresponding to an emission 
measure of 6740 cm$^{-6}$~pc for a temperature of 10$^4$ K.
The ionized gas has a shell structure, and the longest path 
viewed through the shell is along the rim and equals
2$^{3/2}R(\Delta R/R)^{1/2}$, where $R$ and $\Delta R$ are
the shell radius and thickness, respectively.
For a radius of 0.65 pc and a thickness of 0.15 pc (measured
from the H$\alpha$ image), the peak emission measure 
leads to an electron density of $90\pm10$ cm$^{-3}$.
As the temperature of a photoionized gas is of order 
10$^4$ K, the thermal pressure of the ionized skin of the 
dust globule is $P/k \sim 2\times10^6$ cm$^{-3}$ K,
substantially higher than that of N\,51D's interior hot gas.

The thermal pressure in the dust globule is $P/k \sim
10^4$ cm$^{-3}$ K, assuming typical values of density
(10$^3$ H$_2$ cm$^{-3}$) and temperature (10 K) for Bok
globules.  This pressure is much lower than those of the
ionized surfaces or the surrounding hot gas.
The radii of the dust globules are slightly larger than 
the Jeans radius of 0.5 pc, implying that spontaneous
collapse may have started the star formation. 
However, their co-location with the $\sim3$ Myr old OB 
association LH\,54 \citep{OS98} suggests that star formation 
in the globules is delayed and may have been induced recently 
by the thermal pressure of the superbubble interior.
Assuming additional magnetic support, the magnetic pressure
in the dust globules would be $\lesssim$4 times the 
thermal pressure, suggesting a magnetic field of
$\lesssim$35 $\mu$G, compatible with those observed in 
Galactic molecular clouds \citep{CR99}.

\section{Summary}

Using {\it Spitzer} and {\it HST} observations of the superbubble
N\,51D in the LMC, we have identified three YSOs in dust globules.
One of these is associated with an HH object that is detected
outside the Galaxy for the first time.  
The SEDs of these YSOs are consistent with those of early-B stars 
with disk, envelope, and outflow cavities.
The thermal pressure of the dust globule is much lower than 
that of the surrounding warm photoionized gas or the hot 
shock-heated gas in the superbubble interior.
If star formation is induced by the thermal pressure of the 
hot gas, the magnetic field of the dust globule is $\lesssim$35
$\mu$G.

\acknowledgments 
This work is supported by the grant JPL-1264494.


\begin{deluxetable}{cccc}
\tablewidth{0pt}
\tablecaption{Spectral Energy Distribution of Three YSOs in 
N\,51D\tablenotemark{a}}
\tablehead{
                 &  Source 1   &  Source 2   &  Source 3 \\
Wavelength       &  Flux       &  Flux       &  Flux     \\
($\mu$m)         &  (mJy)      &  (mJy)      &  (mJy) 
}
\startdata
~~3.6          & ~~~4.16$\pm$~10\%  &  ~~~1.51$\pm$~10\%  & ~~2.14$\pm$~10\% \\
~~4.5          & ~~~9.35$\pm$~10\%  &  ~~~1.44$\pm$~10\%  & ~~2.35$\pm$~10\% \\
~~5.8          & ~~24.7~$\pm$~10\%  &  ~~~4.71$\pm$~10\%  & ~~5.35$\pm$~10\% \\
~~8.0          & ~~47.6~$\pm$~10\%  &  ~~11.4~$\pm$~10\%  & ~13.3~$\pm$~10\% \\
~24~~          & ~356~~~$\pm$10\%  &  ~~43~~~$\pm$10\%  & ~20~~~$\pm$10\% \\
~70~~          & 1508~~~$\pm$20\%  &  ~429~~~$\pm$20\%  & 180~~~$\pm$20\% \\
160~~          & 1402~~~$\pm$20\%  &  $<$193            &  $<$193
\enddata
\tablenotetext{a}{The errors are dominated by uncertainties in the flux 
calibration, as given in the IRAC and MIPS Data Handbooks. The measurement 
errors are only 2--3\% for sources 1 and 3, and 4--7\% for source 2. }
\end{deluxetable}

\newpage

\begin{figure}
\epsscale{0.5}
\vskip 10cm
\caption{{\it (a)} A color composite of N\,51D with H$\alpha$ in green, 
green continuum in blue, and IRAC 8 $\mu$m in red.  
The H$\alpha$ and green continuum images are from the Magellanic 
Cloud Emission-Line Survey \citep{Setal99}.  
{\it (b)} An IRAC color composite of N\,51D with 3.6 $\mu$m in blue,
4.5 $\mu$m in green, and 8.0 $\mu$m in red.}
\epsscale{1.0}
\end{figure}

\begin{figure}
\plotone{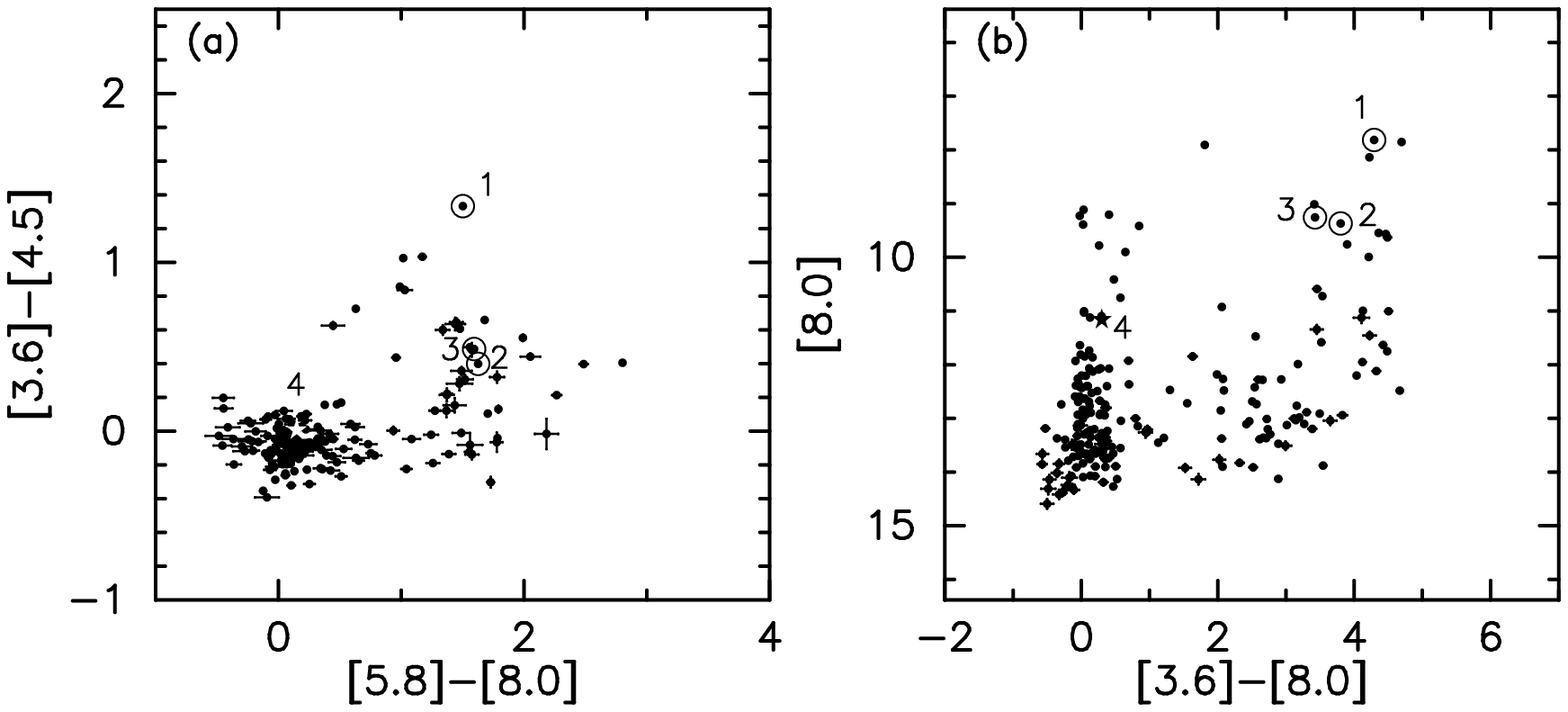}
\caption{{\it (a)} [3.6]$-$[4.5] versus [5.8]$-$[8.0] color-color
diagram of N\,51D.  {\it (b)} [8.0] versus [3.6]$-$[8.0] color-magnitude 
diagram of N\,51D.  Objects 1--3 are marked by circles around the
data points, and object 4 is marked by a star symbol.}
\end{figure}

\begin{figure}
\caption{{\it (a)} A color composite of {\it HST} and {\it Spitzer}
images of N\,51D with H$\alpha$ in green, [\ion{S}{2}] in blue, and
8.0 $\mu$m in red.  Note that the {\it Spitzer} 8 $\mu$m image has 
a $\sim2''$ resolution so the sources appear as smudges, as opposed 
to sharp point sources in the {\it HST} images.
{\it (b)} {\it HST} WFPC2 H$\alpha$ image of YSO-1 and YSO-2.
{\it (c)} {\it HST} WFPC2 H$\alpha$ image of YSO-3.
{\it (d)} {\it HST} WFPC2 [\ion{S}{2}] image of YSO-3.}
\end{figure}

\begin{figure}
\plotone{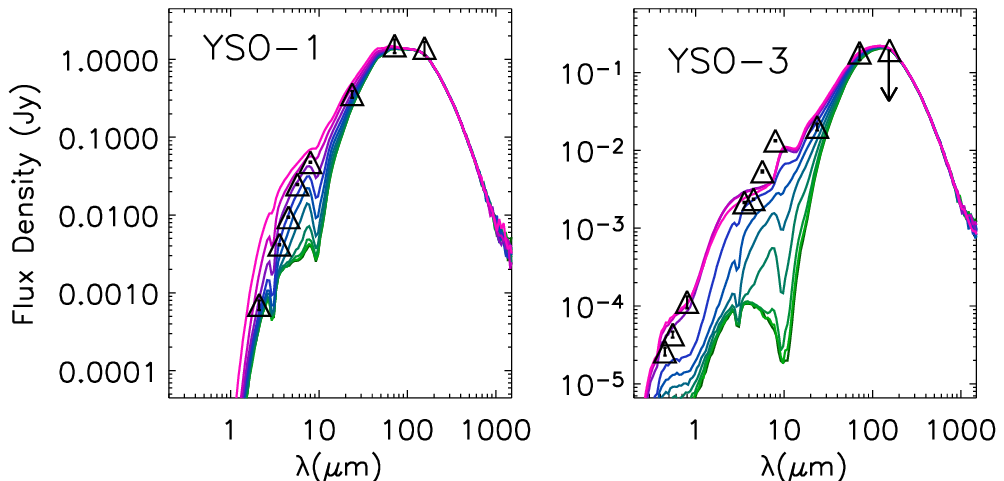}
\caption{Models of SEDs for YSO-1 ({\it left}) and YSO-3 ({\it right}).  
The different colored lines correspond to different inclinations 
ranging from $\cos i = 0.05$ (green) to $0.95$ (pink) in steps 
of 0.1.  The observations are shown as triangles with error bars.}
\end{figure}

\end{document}